\documentclass[12pt]{article}

\author{M.V. Khabarov\thanks{maksim.khabarov@ihep.ru},
Yu.M. Zinoviev\thanks{yurii.zinoviev@ihep.ru}
\\[0.5cm]
\it{\small Institute for High Energy Physics of National
Research Center "Kurchatov Institute"} \\
\it{\small Protvino, Moscow Region, 142281, Russia}}

\title{On massive higher spins in $d=3$}

\date{}

\begin{document}

\maketitle

\begin{abstract}
In this paper we consider a frame-like gauge invariant description of
massive higher spin bosons and fermions in $d=3$ and provide for the
first time a proof that such formulation does describe just one
massive physical degree of freedom with the appropriate helicity. For
this purpose we completely fix all the gauge symmetries and show that
all other auxiliary components vanish on-shell, while the only
remaining highest component satisfies the correct equations. As a
bonus, we show that the Lagrangians for the so-called self-dual
massive spin-3 and spin-4 fields proposed by Aragone and Khoudeir (as
well as their generalization to arbitrary integer and half-integer
spins) can be obtained from the gauge invariant ones by the
appropriate gauge fixing.
\end{abstract}

\thispagestyle{empty}
\newpage
\setcounter{page}{1}

\section{Introduction}

A classical (now called metric-like) formulation for massive higher
spin bosons and fermions has been constructed by Singh-Hagen
\cite{SH74,SH74a}. It requires an introduction of a number of
auxiliary fields which themselves vanish on-shell and provide all
necessary constraints. Later on, a gauge invariant metric-like
formulation for the higher spin massive bosons \cite{Zin01} and
fermions \cite{Met06} were proposed. The main idea was to combine an
appropriate set of massless fields and join them together keeping all
the (modified) gauge symmetries. In this, the description of the
massless fields was based on the Fronsdal formulation
\cite{Fro78,FF78} which uses double traceless completely symmetric
(spin-)tensors, while the gauge transformation parameters are
traceless ones. Taking into account that a double traceless tensor is
equivalent to two traceless ones one can completely fix the gauge
setting half of these traceless components to zero. The result
coincides with the Singh-Hagen formulation.

An alternative (now called frame-like) formulation for the massless
higher spin bosons and fermions was proposed by Vasiliev
\cite{Vas80,LV88,Vas88}. It requires an introduction of the physical
and auxiliary one-forms which enter the free Lagrangian as well as a
set of extra fields which do not enter the free Lagrangian but (being
equivalent to higher derivatives of the physical field) play a very
important role in an interacting theory. A frame-like gauge invariant
formulation for the massive higher spin fields was also constructed
\cite{Zin08b,PV10,KhZ19} along the same lines as in the metric-like
case.

As is well known, in three dimensions all massless higher spin fields
with $s \ge 3/2$ do not have any physical degrees of freedom being a
pure gauge, while an irreducible representation for the massive case
corresponds to just one physical degree of freedom with a helicity 
$+s$ or $-s$. A first formulation for the arbitrary spin massive
bosons and fermions was proposed by Tyutin and Vasiliev \cite{TV97}.
It used a set of Lagrangian multipliers to achieve all necessary
constraints. One more possibility (specific to three dimensions) is a
so-called topologically massive formulation (see e.g. \cite{KP18} and
references therein). At last, a frame-like gauge invariant
formulation for the massive higher spins has been constructed along
the same line as in $d \ge 4$ spaces (see review \cite{BSZ17a} and
references therein). In spite of the fact that massless higher spins
in $d=3$ do not have any physical degrees of freedom, the construction
works and already found some successful applications
\cite{BSZ17a,Zin21}. However, a strict proof that such formulation
does describe just one degree of freedom with the correct helicity
was absent up to now. Our aim here is to fill this gap.

Our strategy here is to fix the gauge and show that all the auxiliary
components vanish on-shell, while the remaining main field does
satisfy the correct equations. Note, that starting with the spin-2 one
faces a situation where one and the same field plays double role being
a gauge field for one transformations and a Stueckelberg field for
another one. So to correctly fix the gauge we decompose our one-forms
into irreducible components and use a metric-like multispinor
formalism
similar to the one used in \cite{KP18}. Then we show that the
equations for the Stueckelberg components follow from the equations
for the other components and thus we can set them to zero directly in
the Lagrangian (see e.g. discussion on this theme in Appendix A of
\cite{BHKMS16}). After that we show that all other auxiliary
components indeed vanish on-shell leaving us with just one field
$\omega^{\alpha(2s)}$ which satisfies
$$
\frac{1}{s} D^\alpha{}_\beta \omega^{\alpha(2s-1)\beta} - M
\omega^{\alpha(2s)} \approx 0, \qquad D_{\beta(2)}
\omega^{\alpha(2s-2)\beta(2)} \approx 0
$$

In the paper \cite{AK93} Aragone and Khoudeir proposed the Lagrangians
for the so-called self-dual massive spin-3 and spin-4 fields
generalizing the massive spin-2 case \cite{AK86}. Such formulation
uses a one-form as the main gauge field as well as a set of zero-forms
as the auxiliary ones. Our second aim in this work is to show that
these Lagrangians as well as their generalization for the arbitrary
spin bosons and fermions can be directly obtained from our gauge
invariant formulation by the appropriate gauge fixing.

\noindent
{\bf Notation and conventions} For simplicity we work in flat three
dimensional space. All objects are one-forms or zero-forms having a
number of completely symmetric local spinor indices which we denote 
$\alpha(n) = (\alpha_1\alpha_2\dots \alpha_n)$. A coordinate
independent description of the background space is provided by the
frame one-form $e^{\alpha(2)}$ and a Lorentz covariant derivative $D$
such that $D \wedge D = 0$. Also we use two and three-forms defined
as:
$$
e^{\alpha(2)} \wedge e^{\beta(2)} = \varepsilon^{\alpha\beta}
E^{\alpha\beta}, \qquad E^{\alpha(2)} \wedge e^{\beta(2)} =
\varepsilon^{\alpha\beta} \varepsilon^{\alpha\beta} E
$$
In the main text wedge product sign will be omitted.

\section{Bosons}

In three dimensions a massive higher spin boson describes in general
two physical degrees of freedom. But in the frame-like formalism it is
possible to separate a complete system into two independent subsystems
each describing just one helicity (see \cite{Zin21} for details).
Such description uses a set of one-forms $\Omega^{\alpha(2k)}$,
$1 \le k \le s-1$ and one zero-form $B^{\alpha(2)}$. The Lagrangian:
\begin{eqnarray}
{\cal L} &=& \sum_{k=1} (-1)^{k+1} [ \frac{1}{2} 
\Omega_{\alpha(2k)} D \Omega^{\alpha(2k)} +
\frac{Ms}{2(k+1)} \Omega_{\alpha(2k-1)\beta} e^\beta{}_\gamma
\Omega^{\alpha(2k-1)\gamma} \nonumber \\
 && \qquad \qquad - a_k \Omega_{\alpha(2k)\beta(2)}
e^{\beta(2)} \Omega^{\alpha(2k)} ] \nonumber \\
 && - B_{\alpha\beta} E^\beta{}_\gamma D B^{\alpha\gamma}
  + Ms E B_{\alpha(2)} B^{\alpha(2)}
 - 2a_0 \Omega_{\alpha\beta} E^\beta{}_\gamma B^{\alpha\gamma}, 
\end{eqnarray}
where
\begin{equation}
a_k{}^2 = \frac{(s+k+1)(s-k-1)}{2(k+1)(2k+3)}M^2,
\end{equation}
is invariant under the following gauge transformations:
\begin{eqnarray}
\delta \Omega^{\alpha(2k)} &=& D \eta^{\alpha(2k)} +
a_k e_{\beta(2)} \eta^{\alpha(2k)\beta(2)} +
\frac{a_{k-1}}{k(2k-1)} e^{\alpha(2)} \eta^{\alpha(2k-2)} \nonumber \\
 && + \frac{Ms}{2k(k+1)} e^\alpha{}_\beta 
\eta^{\alpha(2k-1)\beta},  \\
\delta \Omega^{\alpha(2)} &=& D \eta^{\alpha(2)} + a_1
e_{\beta(2)} \eta^{\alpha(2)\beta(2)} + \frac{Ms}{4} 
e^\alpha{}_\beta \eta^{\alpha\beta}, \nonumber \\
\delta B^{\alpha(2)} &=& 2a_0 \eta^{\alpha(2)}. \nonumber
\end{eqnarray}
Let us consider $\eta^{\alpha(2)}$-transformations:
\begin{equation}
\delta \Omega^{\alpha(2)} = D \eta^{\alpha(2)} + \frac{Ms}{4} 
e^\alpha{}_\beta \eta^{\alpha\beta}, \qquad
\delta \Omega^{\alpha(4)} = \frac{a_1}{6} e^{\alpha(2)}
\eta^{\alpha(2)}, \qquad \delta B^{\alpha(2)} = 2a_0 \eta^{\alpha(2)}.
\end{equation}
From this relations it follows (as we have explicitly checked):
\begin{equation}
D \frac{\delta {\cal L}}{\delta \Omega^{\alpha(2)}} 
 - \frac{Ms}{4} e^\alpha{}_\beta 
\frac{\delta {\cal L}}{\delta \Omega^{\alpha\beta}} 
 - \frac{a_1}{6} e^{\alpha(2)} 
\frac{\delta {\cal L}}{\delta \Omega^{\alpha(4)}}
= 2a_0 \frac{\delta {\cal L}}{\delta B^{\alpha(2)}}.
\end{equation}
Thus the equation for the field $B^{\alpha(2)}$ follows from  the
equations of other fields so we can put $B^{\alpha(2)} = 0$ directly
in the Lagrangian. This gives us the Lagrangian:
\begin{eqnarray}
{\cal L} &=& \sum_{k=1} (-1)^{k+1} [ \frac{1}{2} 
\Omega_{\alpha(2k)} D \Omega^{\alpha(2k)} +
\frac{Ms}{2(k+1)} \Omega_{\alpha(2k-1)\beta} e^\beta{}_\gamma
\Omega^{\alpha(2k-1)\gamma} \nonumber \\
 && \qquad \qquad - a_k \Omega_{\alpha(2k)\beta(2)}
e^{\beta(2)} \Omega^{\alpha(2k)} ],  \label{lag_b}
\end{eqnarray}
which is still invariant under the remaining gauge transformations
$\eta^{\alpha(2k)}$, $k > 1$. Each one-form $\Omega^{\alpha(2k)}$
plays double role being gauge field for one transformation and a 
Stueckelberg field for another one. Thus to correctly fix all the
gauge transformations we use a metric-like multispinor formalism
similar to the one used e.g. in \cite{KP18}. Namely, we decompose each
one-form  into three zero-forms as follows:
\begin{equation}
\Omega^{\alpha(2k)} = e_{\beta(2)} \omega_+^{\alpha(2k)\beta(2)}
+ e^\alpha{}_\beta \omega_0^{\alpha(2k-1)\beta} + e^{\alpha(2)}
\omega_-^{\alpha(2k-2)}, \qquad
D = e^{\alpha(2)} D_{\alpha(2)}.
\end{equation}
Then by a straightforward calculations we  can rewrite the Lagrangian
in terms of these new variables:
\begin{eqnarray}
{\cal L} &=& \sum_{k=1}^{s-1} (-1)^{k+1} [ 
 \omega_{+,\alpha(2k+1)\beta} D^\beta{}_\gamma
\omega_+^{\alpha(2k+1)\gamma} + 2k \omega_{+,\alpha(2k)\beta(2)}
D^{\beta(2)} \omega_0^{\alpha(2k)} \nonumber \\
 && + 2k \omega_{0,\alpha(2k-1)\beta} D^\beta{}_\gamma
\omega_0^{\alpha(2k-1)\gamma} + 2k(k+1)(2k-1)
\omega_{0,\alpha(2k-2)\beta(2)} D^{\beta(2)}
\omega_-^{\alpha(2k-2)} \nonumber \\
 && - k(k-1)(2k-1)(2k+1) \omega_{-,\alpha(2k-3)\beta}
D^\beta{}_\gamma \omega_-^{\alpha(2k-3)\gamma} \nonumber \\
 && + \frac{Ms}{2(k+1)} ( - \omega_{+,\alpha(2k+2)}
\omega_+^{\alpha(2k+2)} + 2(k+1) \omega_{0,\alpha(2k)}
\omega_0^{\alpha(2k)} \nonumber \\
 && \qquad + k(k+1)(2k-1)(2k+1) \omega_{-,\alpha(2k-2)}
\omega_-^{\alpha(2k-2)} ) \nonumber \\
 && \qquad + a_k (  2(k+2) \omega_{0,\alpha(2k+2)}
\omega_+^{\alpha(2k+2)} 
 + 2k(k+1)(2k+3) \omega_{-,\alpha(2k)} \omega_0^{\alpha(2k)}) ].
\end{eqnarray}
The Lagrangian equations for the $\omega_{+0-}$ components (which are
gauge invariant hence notations):
\begin{eqnarray}
{\cal R}_+^{\alpha(2k+2)} &=& \frac{1}{(k+1)} D^\alpha{}_\beta
\omega_+^{\alpha(2k+1)\beta} + \frac{k}{(k+1)(2k+1)}
D^{\alpha(2)} \omega_0^{\alpha(2k)} \nonumber \\
 && - \frac{Ms}{(k+1)} \omega_+^{\alpha(2k+2)} 
+ 2(k+2)a_k \omega_0^{\alpha(2k+2)}, \label{eq+} \\
{\cal R}_0^{\alpha(2k)} &=& - 2k D_{\beta(2)} 
\omega_+^{\alpha(2k)\beta(2)} + 2 D^\alpha{}_\beta
\omega_0^{\alpha(2k-1)\beta} + 2(k+1) D^{\alpha(2)}
\omega_-^{\alpha(2k-2)} \nonumber \\
 && + 2Ms \omega_0^{\alpha(2k)} - 2(k+1)a_{k-1}
\omega_+^{\alpha(2k)} + 2k(k+1)(2k+3)a_k
\omega_-^{\alpha(2k)}, \label{eq0} \\
{\cal R}_-^{\alpha(2k-2)} &=& - 2k(k+1)(2k-1) D_{\beta(2)}
\omega_0^{\alpha(2k-2)\beta(2)} - k(2k-1)(2k+1) D^\alpha{}_\beta
\omega_-^{\alpha(2k-3)\beta} \nonumber \\
 && + k(2k-1)(2k+1)Ms \omega_-^{\alpha(2k-2)} 
- 2k(k-1)(2k+1)a_{k-1} \omega_0^{\alpha(2k-2)}. \label{eq-}
\end{eqnarray}
Now let us consider a concrete gauge transformation 
$\eta^{\alpha(2k)}$:
\begin{eqnarray}
\delta \Omega^{\alpha(2k+2)} &=& \frac{a_k}{(k+1)(2k+1)} e^{\alpha(2)}
\eta^{\alpha(2k)}, \nonumber \\
\delta \Omega^{\alpha(2k)} &=& D \eta^{\alpha(2k)} +
\frac{Ms}{2k(k+1)} e^\alpha{}_\beta \eta^{\alpha(2k-1)\beta}, \\
\delta \Omega^{\alpha(2k-2)} &=& a_{k-1} e_{\beta(2)} 
\eta^{\alpha(2k-2)\beta(2)}. \nonumber
\end{eqnarray}
For the new variables we obtain:
\begin{eqnarray}
\delta \omega_-^{\alpha(2k)} &=& \frac{a_k}{(k+1)(2k+1)}
\eta^{\alpha(2k)}, \nonumber \\
\delta \omega_+^{\alpha(2k+2)} &=& \frac{1}{(k+1)(2k+1)} D^{\alpha(2)}
\eta^{\alpha(2k)}, \nonumber \\
\delta \omega_0^{\alpha(2k)} &=& - \frac{1}{2k(k+1)} D^\alpha{}_\beta
\eta^{\alpha(2k-1)\beta} + \frac{Ms}{2k(k+1)} \eta^{\alpha(2k)}, \\
\delta \omega_-^{\alpha(2k-2)} &=& \frac{1}{k(2k+1)} D_{\beta(2)}
\eta^{\alpha(2k-2)\beta(2)}, \nonumber \\
\delta \omega_+^{\alpha(2k)} &=& a_{k-1} \eta^{\alpha(2k)}. \nonumber
\end{eqnarray}
From these equations we obtain a relation (which we have
explicitly checked):
\begin{eqnarray}
&& - D^\alpha{}_\beta {\cal R}_0^{\alpha(2k-1)\beta} 
- \frac{2(k+1)}{k(2k-1)(2k+1)}
 D^{\alpha(2)} {\cal R}_-^{\alpha(2k-2)} 
+ Ms {\cal R}_0^{\alpha(2k)} 
 - \frac{2ka_k}{(2k+1)} {\cal R}_-^{\alpha(2k)} \nonumber \\
 &=& 2k(k+1)[ D_{\beta(2)} {\cal R}_+^{\alpha(2k)\beta(2)} 
+ a_{k-1} {\cal R}_+^{\alpha(2k)} ]. \label{req_b}
\end{eqnarray}
Taking into account that the highest component $\omega_+^{\alpha(2s)}$
is not a Stueckelberg field, we find that the equation for the
$\omega_+^{\alpha(2s-2)}$ component follows from the equations of
other fields. Then, using (\ref{req_b}) recursively we find that the
equations for all $\omega_+$ components follow from the equations of
other fields. Thus we can completely fix all the gauge symmetries
setting all $\omega_+^{\alpha(2k)} = 0$, $2 \le k \le s-1$.
Then the equations (\ref{eq+}) reduce to
\begin{equation}
{\cal R}_+^{\alpha(2k+2)} = 2(k+2)a_k \omega_0^{\alpha(2k+2)}
+ \frac{k}{(k+1)(2k+1)} D^{\alpha(2)} \omega_0^{\alpha(2k)}
\approx 0, \qquad k \ge 1.
\end{equation}
There is no such equation for $k=0$, but we still can use the equation
(\ref{req_b}) for $k=1$  and obtain:
\begin{equation}
- D^\alpha{}_\beta {\cal R}^{\alpha\beta} - \frac{4}{3} D^{\alpha(2)}
{\cal R}_- + Ms {\cal R}_0^{\alpha(2)} - \frac{2a_1}{3} 
{\cal R}_-^{\alpha(2)} - 4 D_{\beta(2)} {\cal R}_+^{\alpha(2)\beta(2)}
= 16a_0{}^2 \omega_0^{\alpha(2)} \approx 0.
\end{equation}
Thus we see that all components $\omega_0 \approx 0$. At last, from
\begin{eqnarray}
{\cal R}_- &=& 3Ms \omega_- \approx 0, \\
{\cal R}_0^{\alpha(2k)} &=& 2k(k+1)(2k+3)a_k \omega_-^{\alpha(2k)}
+ 2(k+1) D^{\alpha(2)} \omega_-^{\alpha(2k-2)} \approx 0. \nonumber
\end{eqnarray}
we find that all components $\omega_- \approx 0$ as well. This leaves
us with just one main field $\omega_+^{\alpha(2s)}$ satisfying the two
equations:
\begin{equation}
\frac{1}{s} D^\alpha{}_\beta \omega_+^{\alpha(2s-1)\beta}
- M \omega_+^{\alpha(2s)} \approx 0, \qquad
D_{\beta(2)} \omega_+^{\alpha(2s-2)\beta(2)} \approx 0.
\end{equation}

\section{Fermions}

A frame-like description of the massive spin-(s+1/2) fermions is
similar to the one for massive bosons\footnote{Let us stress that for
all the formulas that follow it is important to consider all fermionic
objects (both fields and gauge parameters) as anticommuting ones.}. It
requires a set of one-forms $\Phi^{\alpha(2k+1)}$, $0 \le k \le s-1$
and one zero-form $\phi^\alpha$. The Lagrangian
\begin{eqnarray}
\frac{1}{i} {\cal L} &=& \sum_{k=0}^{s-1} (-1)^{k+1} 
[ \frac{1}{2} \Phi_{\alpha(2k+1)} D \Phi^{\alpha(2k+1)}
 + \frac{(2s+1)M}{2(2k+3)} \Phi_{\alpha(2k)\beta} e^\beta{}_\gamma
\Phi^{\alpha(2k)\gamma} \nonumber \\
 && \qquad \qquad + a_k \Phi_{\alpha(2k-1)\beta(2)} e^{\beta(2)}
\Phi^{\alpha(2k-1)} ] \nonumber \\
 && + \frac{1}{2} \phi_\alpha E^\alpha{}_\beta D \phi^\beta
 - \frac{(2s+1)M}{2} E \phi_\alpha \phi^\alpha 
+ a_0 \Phi_\alpha E^\alpha{}_\beta \phi^\beta,
\end{eqnarray}
where
\begin{equation}
a_k{}^2 = \frac{(s+k+1)(s-k)}{2(k+1)(2k+1)} M^2, \qquad
a_0{}^2 = 2s(s+1)M^2, 
\end{equation}
is invariant under the following gauge transformations:
\begin{eqnarray}
\delta_0 \Phi^{\alpha(2k+1)} &=& D \zeta^{\alpha(2k+1)} +
\frac{(2s+1)M}{(2k+1)(2k+3)} e^\alpha{}_\beta \zeta^{\alpha(2k)\beta}
\nonumber \\
 && + \frac{a_k}{k(2k+1)} e^{\alpha(2)} \zeta^{\alpha(2k-1)}
+ a_{k+1} e_{\beta(2)} \zeta^{\alpha(2k+1)\beta(2)}, \\
\delta_0 \phi^\alpha &=& a_0 \zeta^\alpha. \nonumber
\end{eqnarray}
First, let us consider $\zeta^\alpha$-transformations:
\begin{equation}
\delta \Phi^\alpha = D \zeta^\alpha + \frac{(2s+1)M}{3} 
e^\alpha{}_\beta \zeta^\beta, \qquad \delta \Phi^{\alpha(3)} =
\frac{a_1}{3} e^{\alpha(2)} \zeta^\alpha, \qquad \delta \phi^\alpha =
a_0 \zeta^\alpha.
\end{equation}
This produces a relation (which we explicitly checked):
\begin{equation}
D \frac{\delta {\cal L}}{\delta \Phi^\alpha}
 - \frac{(2s+1)M}{3} e_\alpha{}^\beta 
\frac{\delta {\cal L}}{\delta \Phi^\beta}
 - \frac{a_1}{3} e^{\alpha(2)} 
\frac{\delta {\cal L}}{\delta \Phi^{\alpha(3)}}
= a_0 \frac{\delta {\cal L}}{\delta \phi^\alpha}.
\end{equation}
Thus the equation for the field $\phi^\alpha$ follows from the
equations for other fields and we can put $\phi^\alpha = 0$ directly
in Lagrangian. We obtain then the Lagrangian:
\begin{eqnarray}
\frac{1}{i} {\cal L} &=& \sum_{k=0}^{s-1} (-1)^{k+1} 
[ \frac{1}{2} \Phi_{\alpha(2k+1)} D \Phi^{\alpha(2k+1)}
 + \frac{(2s+1)M}{2(2k+3)} \Phi_{\alpha(2k)\beta} e^\beta{}_\gamma
\Phi^{\alpha(2k)\gamma} \nonumber \\
 && \qquad \qquad + a_k \Phi_{\alpha(2k-1)\beta(2)} e^{\beta(2)}
\Phi^{\alpha(2k-1)} ], \label{lag_f}
\end{eqnarray}
which is still invariant under the remaining gauge transformations. 
We proceed with the transition to the new variables:
\begin{equation}
\Phi^{\alpha(2k+1)} = e_{\beta(2)} \phi_+^{\alpha(2k+1)\beta(2)}
+ e^\alpha{}_\beta \phi_0^{\alpha(2k)\beta} + e^{\alpha(2)}
\phi_-^{\alpha(2k-1)}.
\end{equation}
The Lagrangian in terms of these new variables has the form:
\begin{eqnarray}
\frac{1}{i} {\cal L} &=& \sum_{k=0}^{s-1} (-1)^{k+1} 
[ \phi_{+,\alpha(2k+2)\beta} D^\beta{}_\gamma
\phi_+^{\alpha(2k+2)\gamma} + (2k+1) \phi_{+,\alpha(2k+1)\beta(2)}
D^{\beta(2)} \phi_0^{\alpha(2k+1)} \nonumber  \\
 && + (2k+1) \phi_{0,\alpha(2k)\beta} D^\beta{}_\gamma
\phi_0^{\alpha(2k)\gamma} + k(2k+1)(2k+3) 
\phi_{0,\alpha(2k-1)\beta(2)} D^{\beta(2)} \phi_-^{\alpha(2k-1)}
\nonumber \\
 && - k(k+1)(2k-1)(2k+1) \phi_{-,\alpha(2k-2)\beta}
D^\beta{}_\gamma \phi_-^{\alpha(2k-2)\gamma} \nonumber \\
 && + \frac{(2s+1)M}{2(2k+3)} ( - \phi_{+,\alpha(2k+3)}
\phi_+^{\alpha(2k+3)} + (2k+3) 
 \phi_{0,\alpha(2k+1)} \phi_0^{\alpha(2k+1)} \nonumber \\
 && + k(k+1)(2k+1)(2k+3) \phi_{-,\alpha(2k-1)} \phi_-^{\alpha(2k-1)})
\nonumber \\
 && - a_k ( (2k+3) \phi_{0,\alpha(2k+1)} \phi_+^{\alpha(2k+1)} +
 (k+1)(2k-1)(2k+1) \phi_{-,\alpha(2k-1)} \phi_0^{\alpha(2k-1)} ) ]
\end{eqnarray}
Lagrangian equations for these variables look like:
\begin{eqnarray}
{\cal F}_+^{\alpha(2k+3)} &=& \frac{2}{(2k+3)} D^\alpha{}_\beta
\phi_+^{\alpha(2k+2)\beta} + \frac{(2k+1)}{(k+1)(2k+3)}
D^{\alpha(2)} \phi_0^{\alpha(2k+1)} \nonumber \\
 && - \frac{(2s+1)M}{(2k+3)} \phi_+^{\alpha(2k+3)} 
+ (2k+5)a_{k+1} \phi_0^{\alpha(2k+3)}, \label{feq+} \\
{\cal F}_0^{\alpha(2k+1)} &=& - (2k+1) D_{\beta(2)}
\phi_+^{\alpha(2k+1)\beta(2)} + 2 D^\alpha{}_\beta
\phi_0^{\alpha(2k)\beta} + (2k+3) D^{\alpha(2)}
\phi_-^{\alpha(2k-1)}  \label{feq0} \\
 && + (2s+1)M \phi_0^{\alpha(2k+1)} 
 - (2k+3)a_k \phi_+^{\alpha(2k+1)} + (k+2)(2k+1)(2k+3)a_{k+1}
\phi_-^{\alpha(2k+1)}, \nonumber \\
{\cal F}_-^{\alpha(2k-1)} &=& - k(2k+1)(2k+3)  D_{\beta(2)}
\phi_0^{\alpha(2k-1)\beta(2)} - 2k(k+1)(2k+1) D^\alpha{}_\beta
\phi_-^{\alpha(2k-2)\beta} \nonumber \\
 && + k(k+1)(2k+1)(2s+1)M \phi_-^{\alpha(2k-1)}
- (k+1)(2k-1)(2k+1)a_k \phi_0^{\alpha(2k-1)}. \label{feq-}
\end{eqnarray}
Let us consider $\zeta^{\alpha(2k+1)}$-transformations:
\begin{eqnarray}
\delta \Phi^{\alpha(2k+3)} &=& \frac{a_{k+1}}{(k+1)(2k+3)}
e^{\alpha(2)} \zeta^{\alpha(2k+1)} \nonumber \\
\delta \Phi^{\alpha(2k+1)} &=& D \zeta^{\alpha(2k+1)} +
\frac{(2s+1)M}{(2k+1)(2k+3)} e^\alpha{}_\beta
\zeta^{\alpha(2k)\beta} \\
\delta \Phi^{\alpha(2k-1)} &=& a_k e_{\beta(2)} 
\zeta^{\alpha(2k-1)\beta(2)}. \nonumber
\end{eqnarray}
In terms of the new variables this gives:
\begin{eqnarray}
\delta \phi_-^{\alpha(2k+1)} &=& \frac{a_{k+1}}{(k+1)(2k+3)}
\zeta^{\alpha(2k+1)}, \nonumber \\
\delta \phi_+^{\alpha(2k+3)} &=& \frac{1}{(k+1)(2k+3)} D^{\alpha(2)}
\zeta^{\alpha(2k+1)}, \nonumber \\
\delta \phi_0^{\alpha(2k+1)} &=& - \frac{2}{(2k+1)(2k+3)} 
D^\alpha{}_\beta \zeta^{\alpha(2k)\beta} 
+ \frac{(2s+1)M}{(2k+1)(2k+3)} \zeta^{\alpha(2k+1)}, \\
\delta \omega_-^{\alpha(2k-1)} &=& \frac{1}{(k+1)(2k+1)}
D_{\beta(2)} \zeta^{\alpha(2k-1)\beta(2)}, \nonumber \\
\delta \phi_+^{\alpha(2k+1)} &=& a_k \zeta^{\alpha(2k+1)}. \nonumber
\end{eqnarray}
Hence we obtain the following relation (which we also explicitly
checked):
\begin{eqnarray}
&& - 2 D^\alpha{}_\beta {\cal F}_0^{\alpha(2k)\beta} 
- \frac{(2k+3)}{k(k+1)(2k+1)} D^{\alpha(2)}
{\cal F}_-^{\alpha(2k-1)} + (2s+1)M {\cal F}_0^{\alpha(2k+1)}
\nonumber \\
 && - \frac{(2k+1)a_{k+1}}{(k+1)} {\cal F}_-^{\alpha(2k+1)} 
 = (2k+1)(2k+3) [ D_{\beta(2)} {\cal F}_+^{\alpha(2k+1)\beta(2)} +
a_k  {\cal F}_+^{\alpha(2k+1)} ]. \label{req_f}
\end{eqnarray}
Here also taking into account that the highest component 
$\phi_+^{\alpha(2s+1)}$ is not a Stueckelberg field we see that
equation for the $\phi_+^{\alpha(2s-1)}$ component follows from the
equations for other fields. Then, using relation (\ref{req_f})
recursively we find the same property for all $\phi_+$ components.
Thus we can completely fix all the gauge transformations setting 
$\phi_+^{\alpha(2k+1)} = 0$, $1 \le k \le s-2$. Then the relation
(\ref{feq+}) reduces to:
\begin{equation}
{\cal F}_+^{\alpha(2k+3)} = - (2k+5)a_{k+1} \phi_0^{\alpha(2k+3)}
+ \frac{(2k+1)}{(k+1)(2k+3)} D^{\alpha(2)} \phi_0^{\alpha(2k+1)}
\approx 0. 
\end{equation}
One more equation we obtain from (\ref{req_f}) at $k=0$:
\begin{equation}
 - 2 D^\alpha{}_\beta {\cal F}_0^\beta + (2s+1)M {\cal F}_0^\alpha
- a_1 {\cal F}_-^\alpha - 3 D_{\beta(2)} {\cal F}_+^{\alpha\beta(2)}
= 9a_0{}^2 \phi_0^\alpha \approx 0.
\end{equation}
Thus all the components $\phi_0^{\alpha(2k+1)} \approx 0$. At last,
using
\begin{eqnarray}
{\cal F}_0^\alpha &=& 6a_1 \phi_-^\alpha \approx 0, \\
{\cal F}_0^{\alpha(2k+1)} &=& (k+2)(2k+1)(2k+3)a_{k+1} 
\phi_-^{\alpha(2k+3)} + (2k+3) D^{\alpha(2)} \phi_-^{\alpha(2k-1)}
\approx 0, \nonumber
\end{eqnarray}
we find  that all the components $\phi_-^{\alpha(2k+1)} \approx 0$ as
well. This leaves us with just one highest field 
$\phi_+^{\alpha(2s+1)}$ satisfying
\begin{equation}
\frac{2}{(2s+1)} D^\alpha{}_\beta \phi_+^{\alpha(2s)\beta} 
- M \phi_+^{\alpha(2s+1)} \approx 0, \qquad
D_{\beta(2)} \phi_+^{\alpha(2s-1)\beta(2)} \approx 0.
\end{equation}

\section{Self-dual fields a la Aragone-Khoudeir}

In this section we show that the self-dual Lagrangians for massive
spin-3 and spin-4, proposed by Aragone and Khoudeir in \cite{AK93},
appear as the gauge fixed versions of our gauge invariant Lagrangians.
For completeness, we begin with the self-dual massive spin-2
\cite{AK86}. \\
{\bf Spin 2} The Lagrangian in terms of the one-form 
$\Omega^{\alpha(2)}$ looks like:
\begin{equation}
{\cal L} = \frac{1}{2} \Omega_{\alpha(2)} D \Omega^{\alpha(2)} +
\frac{M}{2} \Omega_{\alpha\beta} e^\beta{}_\gamma
\Omega^{\alpha\gamma}
\end{equation}
and does not have any gauge symmetries. To show, that it indeed
describes just one physical degree of freedom, we again use new
variables:
\begin{equation}
\Omega^{\alpha(2)} = e_{\beta(2)} \omega_+^{\alpha(2)\beta(2)}
+ e^\alpha{}_\beta \omega_0^{\alpha\beta} + e^{\alpha(2)} \omega_-.
\end{equation}
Lagrangian in new variables has the form:
\begin{eqnarray}
{\cal L} &=& \omega_{+,\alpha(2)\beta} D^\beta{}_\gamma
\omega_+^{\alpha(3)\gamma} + 2 \omega_{+,\alpha(2)\beta(2)}
D^{\beta(2)} \omega_0^{\alpha(2)} + 2 \omega_{0,\alpha\beta}
D^\beta{}_\gamma \omega_0^{\alpha\gamma} + 4 \omega_{0,\alpha(2)}
D^{\alpha(2)} \omega_- \nonumber \\
 && - \frac{M}{2} \omega_{+,\alpha(4)} \omega_+^{\alpha(4)}
+ 2M \omega_{0,\alpha(2)} \omega_0^{\alpha(2)} 
+ 3M \omega_-^2.
\end{eqnarray}
Lagrangian equations for the new fields look like:
\begin{eqnarray}
{\cal R}_+^{\alpha(4)} &=& \frac{1}{2} D^\alpha{}_\beta
\omega_+^{\alpha(3)\beta} + \frac{1}{3} D^{\alpha(2)}
\omega_0^{\alpha(2)} - M \Omega_+^{\alpha(4)}, \nonumber \\
{\cal R}_0^{\alpha(2)} &=& - 2 D_{\beta(2)} 
\omega_+^{\alpha(2)\beta(2)} + 2 D^\alpha{}_\beta
\omega_0^{\alpha\beta} + 4 D^{\alpha(2)} \omega_-
+ 4M \omega_0^{\alpha(2)}, \\
{\cal R}_- &=& - 4 D_{\alpha(2)} \omega_0^{\alpha(2)} 
+ 6M \omega_-. \nonumber
\end{eqnarray}
Now by the straightforward calculations we obtain:
\begin{equation}
D_{\beta(2)} {\cal R}_+^{\alpha(2)\beta(2)} + \frac{1}{4}
D^\alpha{}_\beta {\cal R}_0^{\alpha\beta} + \frac{1}{3}
D^{\alpha(2)} {\cal R}_- \frac{M}{2} {\cal R}_0 
= - 2M^2 \omega_0^{\alpha(2)} \approx 0.
\end{equation}
Moreover, it follows that
\begin{equation}
D_{\alpha(2)} \omega_0^{\alpha(2)} + \frac{1}{4} {\cal R}_-
= \frac{3M}{2} \omega_- \approx 0.
\end{equation}
This leaves us with just one field $\omega_+^{\alpha(4)}$ satisfying
\begin{equation}
\frac{1}{2} D^\alpha{}_\beta \omega_+^{\alpha(3)\beta}
- M \omega_+^{\alpha(4)} \approx 0, \qquad
D_{\beta(2)} \omega_+^{\alpha(2)\beta(2)} \approx 0.
\end{equation}
{\bf Spin 3} In this case our initial Lagrangian and gauge
transformations have the form:
\begin{eqnarray}
{\cal L} &=& - \frac{1}{2} \Omega_{\alpha(4)} D \Omega^{\alpha(4)}
- \frac{M}{2} \Omega_{\alpha(3)\beta} e^\beta{}_\gamma
\Omega^{\alpha(3)\gamma} - \frac{M}{2} \Omega_{\alpha(2)\beta(2)}
e^{\beta(2)} \Omega^{\alpha(2)} \nonumber \\
 && + \frac{1}{2} \Omega_{\alpha(2)} D \Omega^{\alpha(2)}
+ \frac{3M}{4} \Omega_{\alpha\beta} e^\beta{}_\gamma
\Omega^{\alpha\gamma}, 
\end{eqnarray}
\begin{equation}
\delta \Omega^{\alpha(4)} = D \eta^{\alpha(4)} + \frac{M}{4}
e^\alpha{}_\beta \eta^{\alpha(3)\beta}, \qquad
\delta \Omega^{\alpha(2)} = \frac{M}{2} e_{\beta(2)} 
\eta^{\alpha(2)\beta(2)}.
\end{equation}
Now we fix the gauge setting (i.e. setting $\omega_+^{\alpha(4)} = 0$)
\begin{equation}
\Omega^{\alpha(2)} = e^\alpha{}_\beta \omega_0^{\alpha\beta} +
e^{\alpha(2)} \omega_-,
\end{equation}
leaving the highest one-form $\Omega^{\alpha(4)}$ intact. We obtain
\begin{eqnarray}
{\cal L} &=& - \frac{1}{2} \omega_{\alpha(4)} D \Omega^{\alpha(4)}
- \frac{M}{2} \Omega_{\alpha(3)\beta} e^\beta{}_\gamma
\Omega^{\alpha(3)\gamma} + 2M \Omega_{\alpha(2)\beta(2)} E^{\beta(2)}
\omega_0^{\alpha(2)} \nonumber \\
 && + 2 E [ 4 \omega_{0,\alpha\beta} D^\beta{}_\gamma
\omega_0^{\alpha\gamma} + 8 \omega_{0,\alpha(2)} D^{\alpha(2)}
\omega_- + 6M \omega_{0,\alpha(2)} \omega_0^{\alpha(2)} + 9M 
\omega_-^2]. 
\end{eqnarray}
The resulting Lagrangian has the same structure as the corresponding
one in \cite{AK93} (up to transition to multispinor formalism).

\noindent
{\bf Spin 4} In this case we have three one-forms
$\Omega^{\alpha(6)}$, $\Omega^{\alpha(4)}$, $\Omega^{\alpha(2)}$ and
two gauge transformations $\eta^{\alpha(6)}$ and $\eta^{\alpha(4)}$.
We fix the gauge setting
\begin{equation}
\Omega^{\alpha(4)} = e^\alpha{}_\beta \omega_0^{\alpha(3)\beta}
+ e^{\alpha(2)} \omega_-^{\alpha(2)}, \qquad
\Omega^{\alpha(2)} = e^\alpha{}^\beta \omega_0^{\alpha\beta}
+ e^{\alpha(2)} \omega_-,
\end{equation}
leaving the highest one-form $\Omega^{\alpha(6)}$ intact. We obtain
the
Lagrangian in  the form:
\begin{eqnarray}
{\cal L} &=& \frac{1}{2} \Omega_{\alpha(6)} D
\Omega^{\alpha(6)} + \frac{M}{2} \Omega_{\alpha(5)\beta}
e^\beta{}_\gamma \Omega^{\alpha(5)\gamma} - 8a_2
\Omega_{\alpha(4)\beta(2)} E^{\beta(2)} \omega_0^{\alpha(4)} \nonumber
\\
 && + 4E [ - 4 \omega_{0,\alpha(3)\beta} D^\beta{}_\gamma
\omega_0^{\alpha(3)\gamma} - 18 \omega_{0,\alpha(2)\beta(2)}
D^{\beta(2)} \omega_-^{\alpha(2)} + 30 \omega_{-,\alpha\beta}
D^\beta{}_\gamma \omega_-^{\alpha\gamma} \nonumber \\
 && \qquad + 2 \omega_{0,\alpha\beta} D^\beta{}_\gamma
\omega_0^{\alpha\gamma} + 2 \omega_{0,\alpha(2)} D^{\alpha(2)}
\omega_-  + 10a_1 \omega_{0,\alpha(2)} \omega_-^{\alpha(2)} \nonumber 
\\
 && \qquad - 2M ( 2\omega_{0,\alpha(4)}
\omega_0^{\alpha(4)} + 30 \omega_{-,\alpha(2)} \omega_-^{\alpha(2)}
 - 2\omega_{0,\alpha(2)} \omega_0^{\alpha(2)} 
- 3 \omega_- \omega_- ) ],
\end{eqnarray}
which also has the same general structure as the corresponding one in
\cite{AK93}.

Similarly, it is possible to obtain the generalization of such
construction to arbitrary integer spin, though it appears to be rather
complicated. Let us turn to the half-integer spins instead. Again for
completes we begin with the simplest spin-3/2 case. \\
{\bf Spin 3/2} The Lagrangian has a very simple form:
\begin{equation}
{\cal L} = \frac{1}{2} \Phi_\alpha D \Phi^\alpha 
+ \frac{M}{2} \Phi_\alpha e^\alpha{}_\beta \Phi^\beta
\end{equation}
and does not have any gauge symmetries. To show that it indeed
describes just one physical degree of freedom we introduce new
variables:
\begin{equation}
\Phi^\alpha = e_{\beta(2)} \phi^{\alpha\beta(2)} + e^\alpha{}_\beta
\phi^\beta. 
\end{equation}
Lagrangian in terms of the new variables looks like:
\begin{equation}
{\cal L} = \phi_{\alpha(2)\beta} D^\beta{}_\gamma 
\phi^{\alpha(2)\gamma} + \phi_{\alpha(2)\beta} D^{\alpha(2)}
\phi^\beta + \phi_\alpha D^\alpha{}_\beta \phi^\beta
 - \frac{M}{2} \phi_{\alpha(3)} \phi^{\alpha(3)} + \frac{3M}{2}
\phi_\alpha \phi^\alpha,
\end{equation}
while Lagrangian equations have the form:
\begin{eqnarray}
{\cal R}^{\alpha(3)} &=& \frac{2}{3} D^\alpha{}_\beta 
\phi^{\alpha(2)\beta} + \frac{1}{3} D^{\alpha(2)} \phi^\alpha
- M \phi^{\alpha(3)}, \nonumber \\
{\cal R}^\alpha &=& - D_{\beta(2)} \phi^{\alpha\beta(2)} + 2
D^\alpha{}_\beta \phi^\beta + 3M \phi^\alpha.
\end{eqnarray}
By direct and simple calculations we obtain:
\begin{equation}
D_{\beta(2)} {\cal R}^{\alpha\beta(2)} + \frac{2}{3} D^\alpha{}_\beta
{\cal R}^\beta - M {\cal R}^\alpha = - 3M^2 \phi^\alpha \approx 0,
\end{equation}
and as a result:
\begin{equation}
\frac{2}{3} D^\alpha{}_\beta \phi^{\alpha(2)\beta} - M
\phi^{\alpha(3)} \approx 0, \qquad 
D_{\beta(2)} \phi^{\alpha\beta(2)} \approx 0.
\end{equation}
{\bf Spin 5/2} For our last example the initial Lagrangian and gauge
transformations look like:
\begin{eqnarray}
{\cal L} &=& \frac{1}{2} \Phi_{\alpha(3)} D \Phi^{\alpha(3)}
+ \frac{M}{2} \Phi_{\alpha(2)\beta} e^\beta{}_\gamma
\Phi^{\alpha(2)\gamma} + a_1 \Phi_{\alpha\beta(2)} e^{\beta(2)}
\Phi^\alpha \nonumber \\
 && - \frac{1}{2} \Phi_\alpha D \Phi^\alpha - \frac{5M}{6}
\Phi_\alpha e^\alpha{}_\beta \Phi^\beta, \qquad 
a_1{}^2 = \frac{M^2}{3}
\end{eqnarray}
\begin{equation}
\delta \Phi^{\alpha(3)} = D \zeta^{\alpha(3)} + \frac{M}{3} 
e^\alpha{}_\beta \zeta^{\alpha(2)\beta}, \qquad
\delta \Phi^\alpha = a_1 e_{\beta(2)} \zeta^{\alpha\beta(2)},
\end{equation}
Now we fix the gauge setting:
\begin{equation}
\Phi^\alpha = e^\alpha{}_\beta \phi_0^\beta,
\end{equation}
leaving the highest one-form $\Phi^{\alpha(3)}$ intact. 
Then we obtain:
\begin{eqnarray}
{\cal L} &=& \frac{1}{2} \Phi_{\alpha(3)} D \Phi^{\alpha(3)}
+ \frac{M}{2} \Phi_{\alpha(2)\beta} e^\beta{}_\gamma
\Phi^{\alpha(2)\gamma} - 2a_1 \Phi_{\alpha\beta(2)} E^{\beta(2)}
\phi_0^\alpha \nonumber \\
 && - E [ 4 \phi_{0,\alpha} D^\alpha{}_\beta \phi_0^\beta 
 + 10M \phi_{0,\alpha} \phi_0^\alpha ].
\end{eqnarray}
Note that as in the bosonic case it is possible to obtain a
generalization of such construction to the arbitrary half-integer
spin.

\section*{Conclusion}

In this note we have shown that the frame-like gauge invariant
Lagrangians constructed previously do describe just one physical
degree of freedom with the appropriate helicity.  For this purpose we
completely fix all the gauge symmetries and show that all other
auxiliary components vanish on-shell, while the only remaining highest
component satisfies the correct equations. As a bonus, we show that
the Lagrangians for the so-called self-dual massive spin-3 and spin-4
fields proposed by Aragone and Khoudeir (as well as their
generalization to arbitrary integer and half-integer spins) can be
obtained from the gauge invariant ones by the appropriate gauge
fixing. Note however that such description appears to be much more
complicated as the initial gauge invariant one.

We hope that the (relatively) simple gauge invariant Lagrangians
(\ref{lag_b}) and (\ref{lag_f}) can serve as a nice starting point for
the investigation of the Lorentz covariant formulation for the massive
higher spin interactions in $d=3$ (for the light-cone formulation
see \cite{Met20,STT20,Met21}).

\end{document}